\documentstyle[epsfig,12pt]{article}
\input epsf
\newlength{\dinwidth}
\newlength{\dinmargin}
\newcommand {\pom} {I\!\!P}
\newcommand {\pomsub} {{\scriptscriptstyle \pom}}

\newcommand {\xpom} {x_{\pomsub}}

\begin{document}

\def\ap#1#2#3   {{\em Ann. Phys. (NY)} {\bf#1} (#2) #3.}   
\def\apj#1#2#3  {{\em Astrophys. J.} {\bf#1} (#2) #3.} 
\def\apjl#1#2#3 {{\em Astrophys. J. Lett.} {\bf#1} (#2) #3.}
\def\app#1#2#3  {{\em Acta. Phys. Pol.} {\bf#1} (#2) #3.}
\def\ar#1#2#3   {{\em Ann. Rev. Nucl. Part. Sci.} {\bf#1} (#2) #3.}
\def\cpc#1#2#3  {{\em Computer Phys. Comm.} {\bf#1} (#2) #3.}
\def\epj#1#2#3  {{\it Eur. Phys. J.} {\bf#1} (#2) #3}
\def\err#1#2#3  {{\it Erratum} {\bf#1} (#2) #3.}
\def\ib#1#2#3   {{\it ibid.} {\bf#1} (#2) #3.}
\def\jmp#1#2#3  {{\em J. Math. Phys.} {\bf#1} (#2) #3.}
\def\ijmp#1#2#3 {{\em Int. J. Mod. Phys.} {\bf#1} (#2) #3}
\def\jetp#1#2#3 {{\em JETP Lett.} {\bf#1} (#2) #3.}
\def\jpg#1#2#3  {{\em J. Phys. G.} {\bf#1} (#2) #3.}
\def\mpl#1#2#3  {{\em Mod. Phys. Lett.} {\bf#1} (#2) #3.}
\def\nat#1#2#3  {{\em Nature (London)} {\bf#1} (#2) #3.}
\def\nc#1#2#3   {{\em Nuovo Cim.} {\bf#1} (#2) #3.}
\def\nim#1#2#3  {{\em Nucl. Instr. Meth.} {\bf#1} (#2) #3.}
\def\np#1#2#3   {{\em Nucl. Phys.} {\bf#1} (#2) #3}
\def\npps#1#2#3   {{\em Nucl. Phys. Proc. Suppl.} {\bf#1} (#2) #3}
\def\pcps#1#2#3 {{\em Proc. Cam. Phil. Soc.} {\bf#1} (#2) #3.}
\def\pl#1#2#3   {{\em Phys. Lett.} {\bf#1} (#2) #3}
\def\prep#1#2#3 {{\em Phys. Rep.} {\bf#1} (#2) #3.}
\def\prev#1#2#3 {{\em Phys. Rev.} {\bf#1} (#2) #3}
\def\prl#1#2#3  {{\em Phys. Rev. Lett.} {\bf#1} (#2) #3}
\def\prs#1#2#3  {{\em Proc. Roy. Soc.} {\bf#1} (#2) #3.}
\def\ptp#1#2#3  {{\em Prog. Th. Phys.} {\bf#1} (#2) #3.}
\def\ps#1#2#3   {{\em Physica Scripta} {\bf#1} (#2) #3.}
\def\rmp#1#2#3  {{\em Rev. Mod. Phys.} {\bf#1} (#2) #3}
\def\rpp#1#2#3  {{\em Rep. Prog. Phys.} {\bf#1} (#2) #3.}
\def\sjnp#1#2#3 {{\em Sov. J. Nucl. Phys.} {\bf#1} (#2) #3}
\def\spj#1#2#3  {{\em Sov. Phys. JEPT} {\bf#1} (#2) #3}
\def\spu#1#2#3  {{\em Sov. Phys.-Usp.} {\bf#1} (#2) #3.}
\def\zp#1#2#3   {{\em Zeit. Phys.} {\bf#1} (#2) #3}

\title{\vspace*{3cm}
\bf{ The legacy of HERA - the first decade}
\vspace*{2cm}}

\author{
 {\bf Aharon Levy} \\ 
{\small \sl School of Physics and Astronomy}\\ {\small \sl Raymond and 
Beverly Sackler Faculty of Exact Sciences}\\
  {\small \sl Tel--Aviv University, Tel--Aviv, Israel}
}
\date{ }
\maketitle

\vspace{5cm}

\begin{abstract}
The $ep$ HERA collider started operation in summer of 1992. This talk
summarizes some of the highlights of physics results obtained since
then and discusses their impact on our understanding of Quantum
Chromodynamics (QCD).
\end{abstract}

\vspace{-20cm}
\begin{flushleft}
DESY 00-116 \\
August 2000 \\
\end{flushleft}

\setcounter{page}{0}
\thispagestyle{empty}
\newpage

\section{Preamble}

This report is based on a talk, presented at DIS2000 in Liverpool,
with the purpose to summarize shortly the legacy of HERA during the
first decade of its operation. The talk was given to an audience
working in the field. For the benefit of the reader who is less
familiar with the subject, the proceedings version has been extended
to include an appendix with definitions of the kinematical variables
used in the text.

\section{Introduction}

The main motivation for building the HERA $ep$ collider was to
continue and study the inner structure of matter and the nature of
forces in a way similar to that of Rutherford, by probing the proton
with higher and higher gauge-boson virtuality, $Q^2$, in order to resolve
smaller and smaller distances. This way of probing the proton is known
as deep inelastic scattering (DIS).

\begin{figure}[h]
\centerline{\psfig{figure=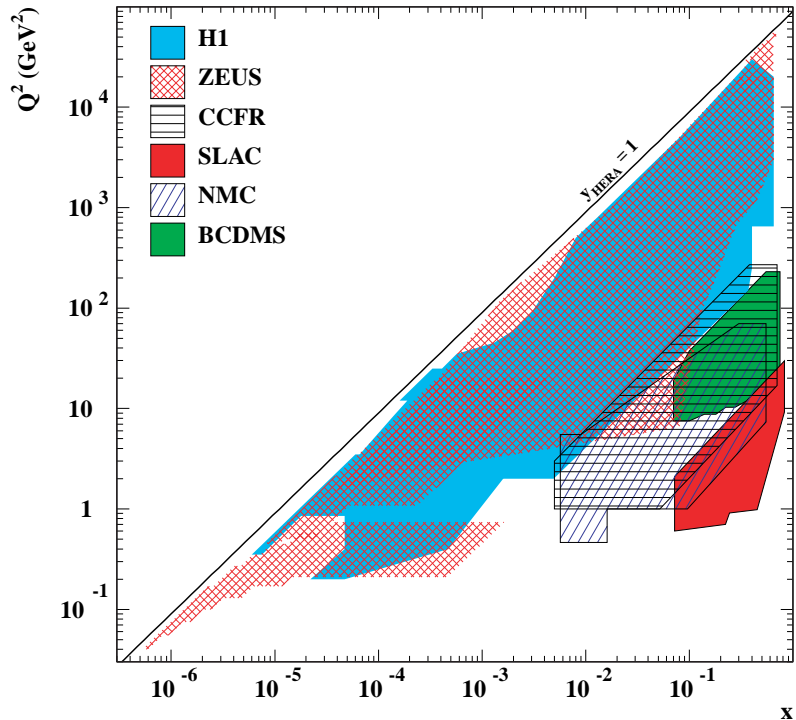,height=10cm}}
\caption{
The $x$-$Q^2$ kinematic plane of some of the fixed target and of the
HERA collider DIS experiments.}
\label{fig:x-q2}
\end{figure}

The plane showing the fraction of the proton's momentum carried by the 
probed parton, $x$, and the virtuality $Q^2$ of the probing photon,
is shown in figure~\ref{fig:x-q2} and is impressive. It
shows how HERA has extended the kinematic reach in both direction in
$Q^2$, low and high, and in the low $x$ region, by few orders of magnitude.

\section{The rise of $F_2$ with decreasing $x$}

\begin{figure}[h]
\centerline{
\psfig{figure=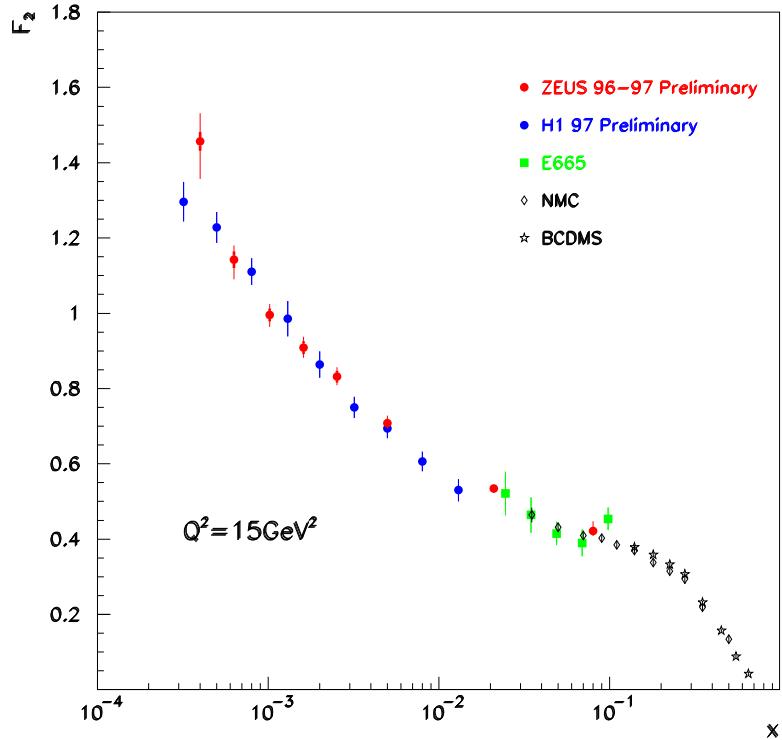,height=10cm}}
\caption{
The proton structure function $F_2$ as function of $x$ at a fixed
value of $Q^2$=15GeV$^2$.
}
\label{fig:f2-q2-15}
\end{figure}

Before this talk, I polled the view of several eminent people as to what they
consider to be the highlights of HERA so far. All of them unanimously
put as number one the surprising sharp rise of $F_2$ with decreasing
$x$~\cite{f2x}, an example of which is shown in figure~\ref{fig:f2-q2-15}.
The DGLAP evolution equations predict the rise of the parton
distributions with decreasing $x$. Why is then the observed rise so
surprising? The physical quantity which is measured is $F_2$. To
decompose $F_2$ into quarks, we need the `impulse approximation' to be
valid.  This means that the configuration of the Fock states into
which the proton fluctuates has to be 'frozen' during the interaction
time. This condition leads~\cite{impulse} to a relation between the
$Q^2$ of the probing photon and some other scales like quark masses,
$m_q$, quark transverse momentum $k_T$'s and $x$. The fluctuation
time, $\tau_f$, is inversely proportional to the mass squared of the
probed configuration, $\tau_f \sim 1/M^2$. The interaction time,
$\tau_{int}$, is inversely proportional to the photon virtuality,
$\tau_{int} \sim 1/Q^2$. The 'freezing' condition for the use of
impulse approximation requires $\tau_f \gg \tau_{int}$ or,
\begin{equation}
Q^2 \gg \sum_i\frac{(m_q^2 + k_{iT}^2)}{x_i}.
\end{equation}
It is therefore highly non trivial that at very low $x$ the
expectations of the DGLAP evolution equations, which apply to partons, 
remain valid for $F_2$. However, HERA teaches us that at 
$x \sim 10^{-3}$ the impulse approximation may be valid down to
$Q^2 \sim$ 1.5 GeV$^2$.
As a summary of this point I
would like to quote Peter Landshoff: 
{\it The amazing discovery of HERA is the rapid rise 
of $F_2$ at small $x$. This remains a challenge for theory: there is
no respectable theoretical understanding.}

\section{Diffraction in DIS}

The second subject, unanimously acclaimed as highlight,
is the observation of abundant
diffraction in DIS~\cite{diffz,diffh1}. This came as a surprise
because if $F_2$ is an incoherent sum of quarks, a DIS process on a
quark can not be diffractive because of color. A diffractive DIS
reaction could happen only if the Fock state of the proton at the
origin includes colorless objects. 

\begin{figure}[h]
\centerline{
\psfig{figure=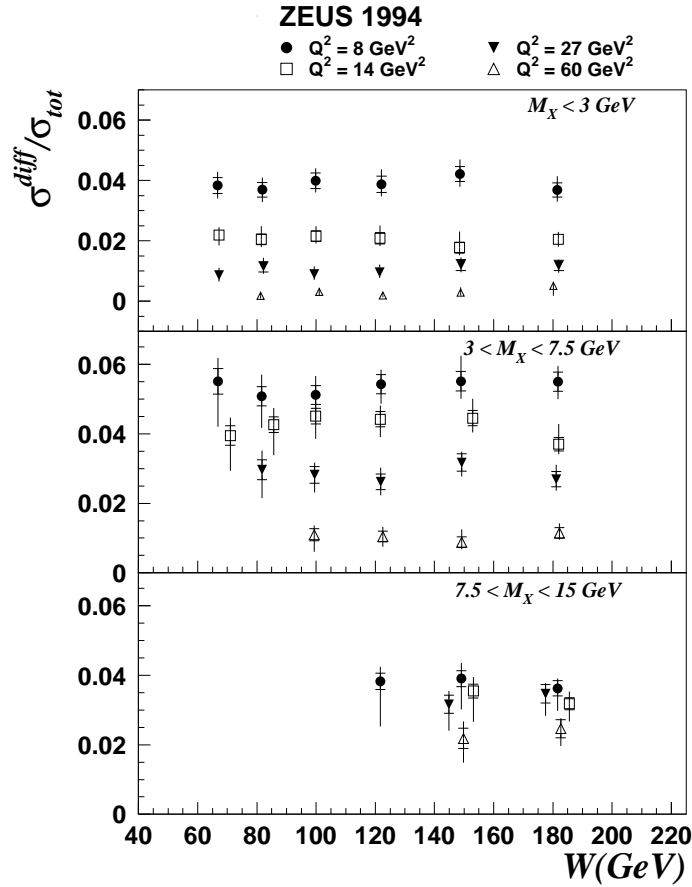,height=12cm}}
\caption{
The ratio of the diffractive to total DIS cross sections as function
of $W$, for given kinematical cuts, as indicated in the figure.
}
\label{fig:diff/tot}
\end{figure}

If this were indeed the case, one would expect that
diffractive DIS processes would be soft in nature, contrary to the
accumulated evidence.
One of the most striking behaviour of these large
rapidity gap events is that they have a similar energy behaviour as
that of the total inclusive cross section~\cite{diffz}, as shown in
figure~\ref{fig:diff/tot},  
not expected in soft processes. Does that mean
that these diffractive DIS processes are all hard? 

\section{Soft-hard interplay at low $x$}

When one thinks about a DIS process involving partons, the first
intuition is to think of it as a hard process. This however turns out
not to be the case. The best way to understand this is when the
interaction is viewed in the proton rest frame. The photon fluctuates
into a $q\bar{q}$ pair, which can either be a large spatial
configuration, or a small one.  The aligned jet model of
Bjorken~\cite{ajm}, leading to scaling properties of $F_2$, assumes
that only large configurations contribute to DIS cross section. That
would make DIS a predominantly soft process. The small configurations
lead to scaling violation, as expected in QCD. To the extend to which
scaling violation introduces only logarithmic corrections to scaling,
one is led to conclude that, even within QCD, DIS is indeed
predominantly soft. To me, it came as a cultural shocks to learn that
DIS does not necessarily mean a hard process.

\subsection{Large and small configurations}

To confirm the validity of the picture above,
we would like to isolate events in which we know that the photon
predominantly fluctuates into a small configuration. This is achieved
either by studying events initiated by longitudinal photons
($\gamma^*_L$)  and
looking at the behaviour of $\sigma_L$, or by studying exclusive
vector meson electroproduction and deeply virtual Compton scattering
(DVCS). 

\begin{figure}[h]
\centerline{
\psfig{figure=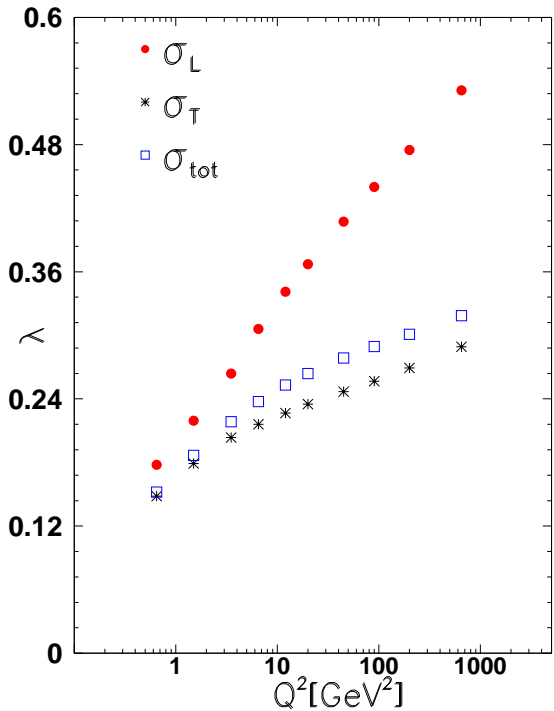,height=10cm}}
\caption{
The dependence of the $W$ exponent $\lambda$ on $Q^2$, for the total
$\gamma^* p$ cross section, $\sigma_{tot}$, the longitudinal one,
$\sigma_L$ and the transverse one, $\sigma_T$. 
}
\label{fig:sigl}
\end{figure}
Since, unfortunately, we did not measure the ratio
$R=\sigma_L/\sigma_T$ of the longitudinal to transverse cross sections
at HERA, I used as an exercise the parameterization of $R$ by Badelek,
Kwiecinski and Stasto~\cite{bks} together with that of
ALLM97~\cite{allm97} as a good representation of
$\sigma_{tot}(\gamma^* p)$, to calculate the effective $\gamma^* p$
energy dependence of the cross section of $\sigma_L$ and $\sigma_T$
separately, assuming a $W^\lambda$ dependence.  The result is shown in
figure~\ref{fig:sigl}. 
As one sees, $\sigma_L$ has a very steep rise
with $Q^2$, as expected from a small configuration process, reaching
the value of 0.5 at $Q^2 \sim$ 1000 GeV$^2$.

\begin{figure}[h]
\centerline{
\psfig{figure=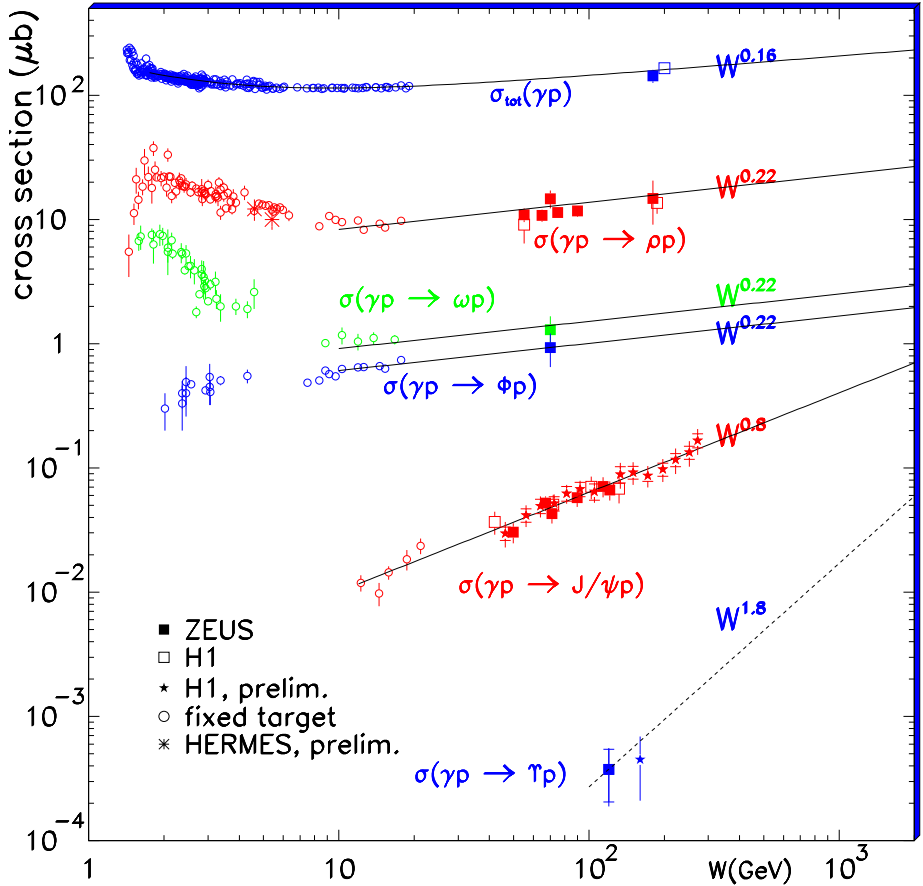,height=13cm}}
\caption{
The $W$ dependence of the total photoproduction cross section together 
with the cross section for photoproduction of vector mesons.
}
\label{fig:xsect-vm}
\end{figure}

\begin{figure}[h]
\centerline{
\psfig{figure=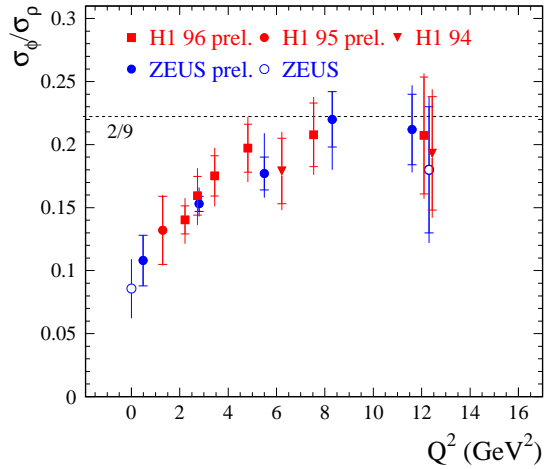,height=8cm}}
\vspace*{-1cm}
\caption{
The ratio of the cross sections of $\phi$ and $\rho^0$ as function of $Q^2$.
}
\label{fig:phi/rho}
\end{figure}

The energy dependence of the cross section for photoproduction of
vector mesons~\cite{ha}, shown in figure~\ref{fig:xsect-vm}, indeed
confirmed that when the photon is squeezed into a small size, there is
a definite change in the $W$ behaviour, as expected from pQCD, which
connects the $W$ behaviour with the rise of the gluons. 
In the case of
$J/\psi$ this happens already at $Q^2$=0, due to the presence of the
charm quarks. For the lighter vector mesons, it happens around
$Q^2$=10 GeV$^2$, as seen from the $\phi/\rho^0$ ratio~\cite{ha},
shown in figure~\ref{fig:phi/rho}, and which reaches the expected
value of 2/9 at that $Q^2$.

\begin{figure}[h]
\vspace*{0.25cm}
\centerline{
\psfig{figure=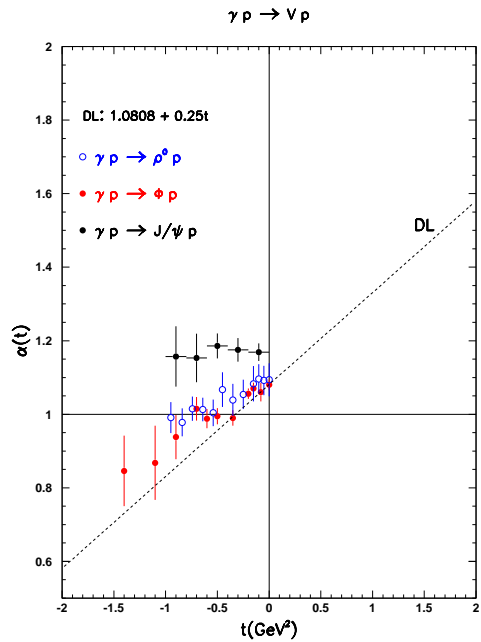,height=8cm}}
\caption{
The Pomeron trajectory as obtained from elastic photoproduction of
$\rho^0$, $\phi$ and $J/\psi$.
}
\label{fig:a-vector}
\end{figure}

The other property of hard exclusive processes is the universal slope
of the diffractive peak. In Regge language this means a decreasing
slope $\alpha^\prime$ of the Regge trajectory. This is shown in
figure~\ref{fig:a-vector} where the Pomeron trajectory is presented as
obtained from elastic photoproduction of $\rho^0$, $\phi$ and
$J/\psi$~\cite{ha}. 
One sees a clear flattening in case of the
$J/\psi$.

\subsection{The gluon and saturation?}

\begin{figure}[h]
\begin{minipage}{6cm}
\centerline{
\psfig{figure=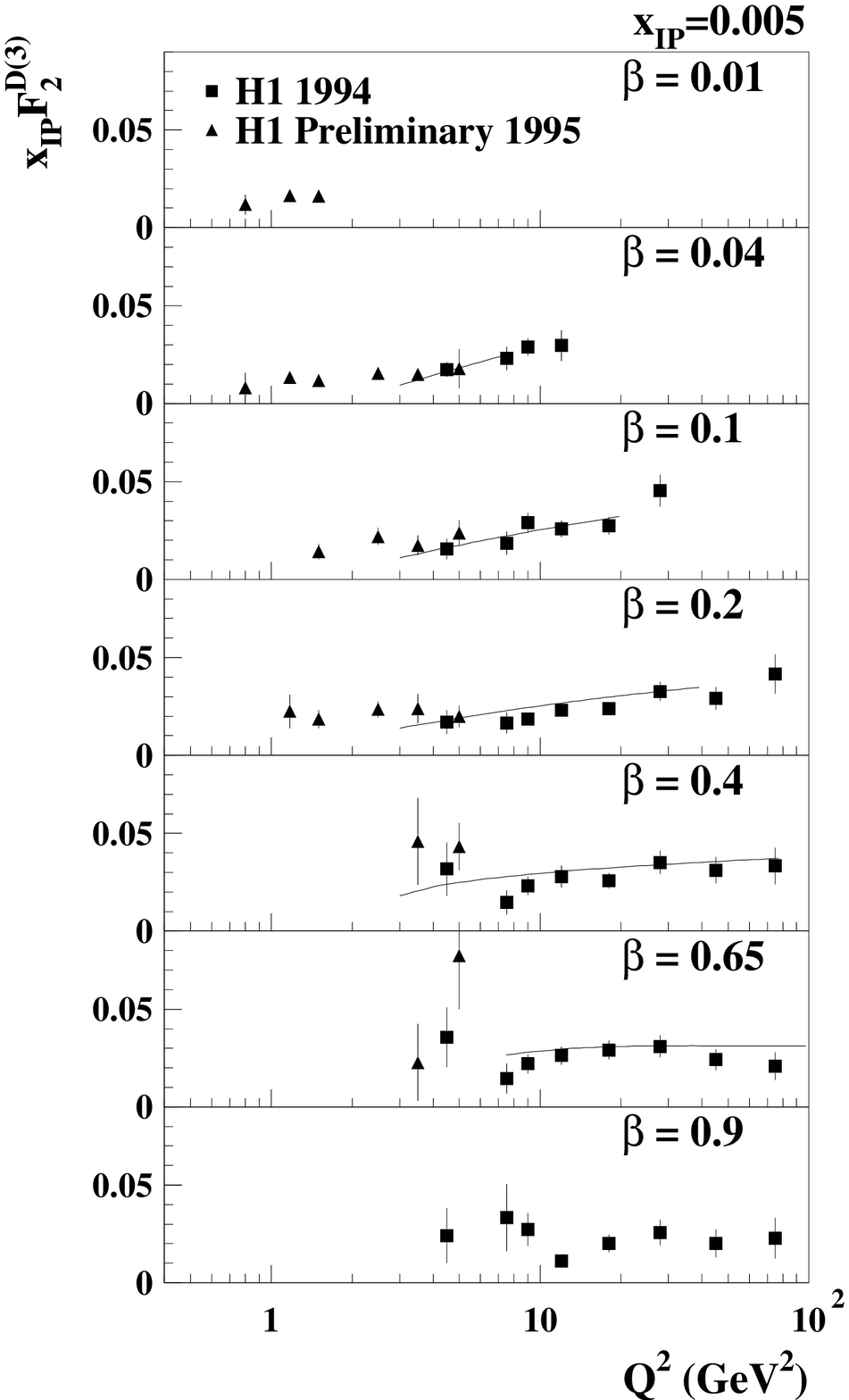,height=9cm}}
\caption{
The inclusive diffractive structure function as function of $Q^2$ for
different fixed values of $\beta$, where $\beta$ is the equivalent of
Bjorken $x$, relative to the colorless exchange.
}
\label{fig:diff}
\end{minipage}
\hspace*{1mm}
\begin{minipage}{10.7cm}
\centerline{
\psfig{figure=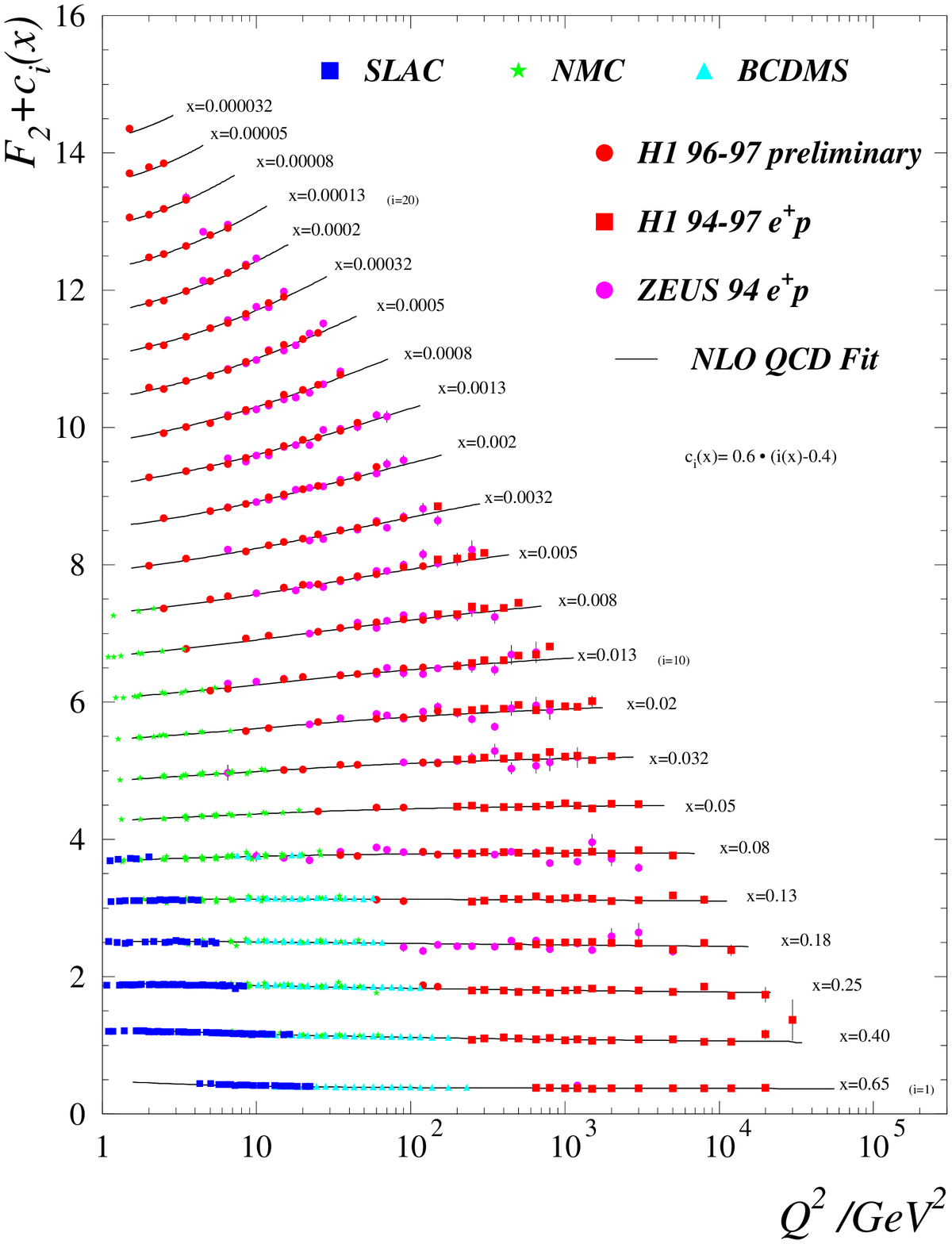,height=10cm}}
\caption{
The proton structure function $F_2$ as function of $Q^2$, for
different fixed values of $x$.
}
\label{fig:sc-viol}
\end{minipage}
\end{figure}

The inclusive diffraction structure function~\cite{diffh1}, presented
in figure~\ref{fig:diff} shows a very different behaviour from that of
the proton structure function~\cite{max}, shown in
figure~\ref{fig:sc-viol}.
While the proton structure function shows positive scaling violation
at low $x$ and negative one at high $x$, the diffractive structure
function shows practically only positive scaling violation, and may
reach scaling at a large values of $\beta$.
QCD factorization was proven to hold also in the inclusive diffraction
case and thus DGLAP evolution equations could be applied. However it
turned out that for the evolution equations to be able to describe the
$Q^2$ behaviour of the data, a large diffractive gluon component is
required~\cite{diffh1}.

\begin{figure}[h]
\begin{minipage}{8.2cm}
\centerline{
\psfig{figure=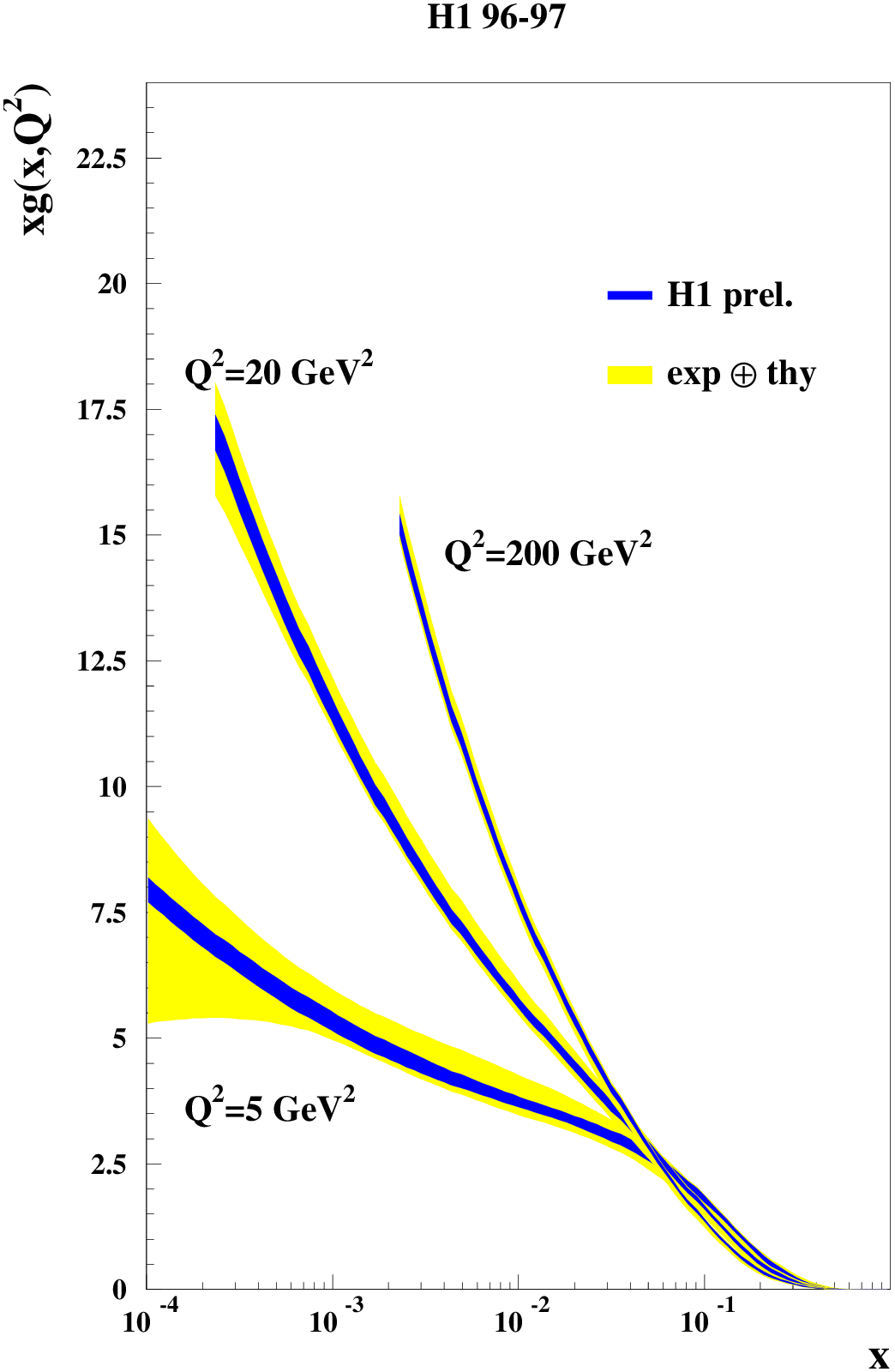,height=8.0cm}}
\caption{
The gluon momentum density distribution as function of $x$ for
different fixed values of $Q^2$.
}
\label{fig:gluon-h1}
\end{minipage}
\hspace*{2mm}
\begin{minipage}{8.2cm}
\centerline{
\psfig{figure=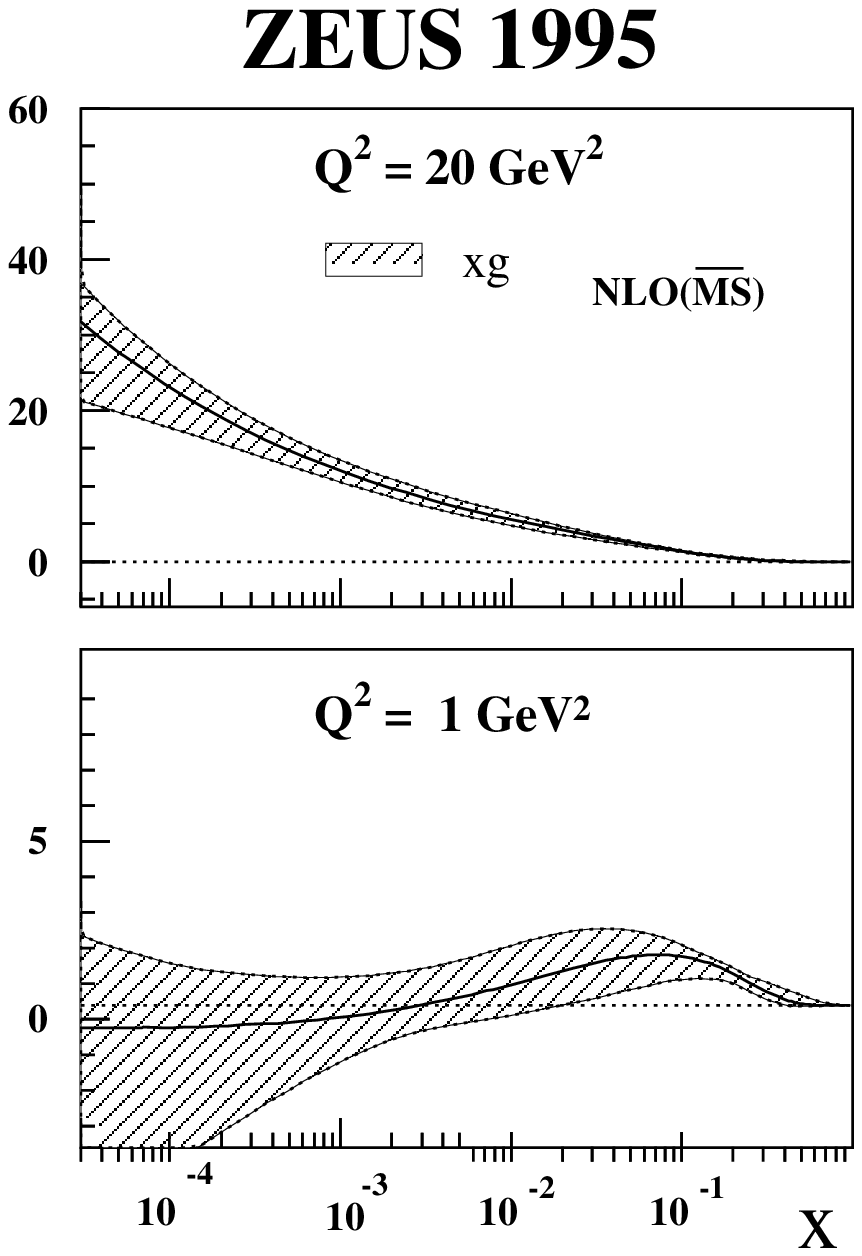,height=8cm}}
\caption{
The gluon momentum density distribution as function of $x$ for
different fixed values of $Q^2$.
}
\label{fig:gluon-z}
\end{minipage}
\end{figure}

This brings me to talk about the gluon in the proton. What have we
learned about the gluon? From the scaling violation, using the
impressive measurements at HERA, shown in figure~\ref{fig:sc-viol},
the precision of the gluon determination has improved tremendously, as
shown in figure~\ref{fig:gluon-h1}~\cite{max}.  The direct gluon
determination from charm production is in very good agreement with the
indirect ones from scaling violation. However, when trying to stretch
the evolution equations down to low $Q^2$, as presented in
figure~\ref{fig:gluon-z} for $Q^2$ = 1 GeV$^2$, one finds a strange
behaviour of the gluon at low $x$~\cite{lowq2z}.

\begin{figure}[h]
\begin{minipage}{9.2cm}
\psfig{figure=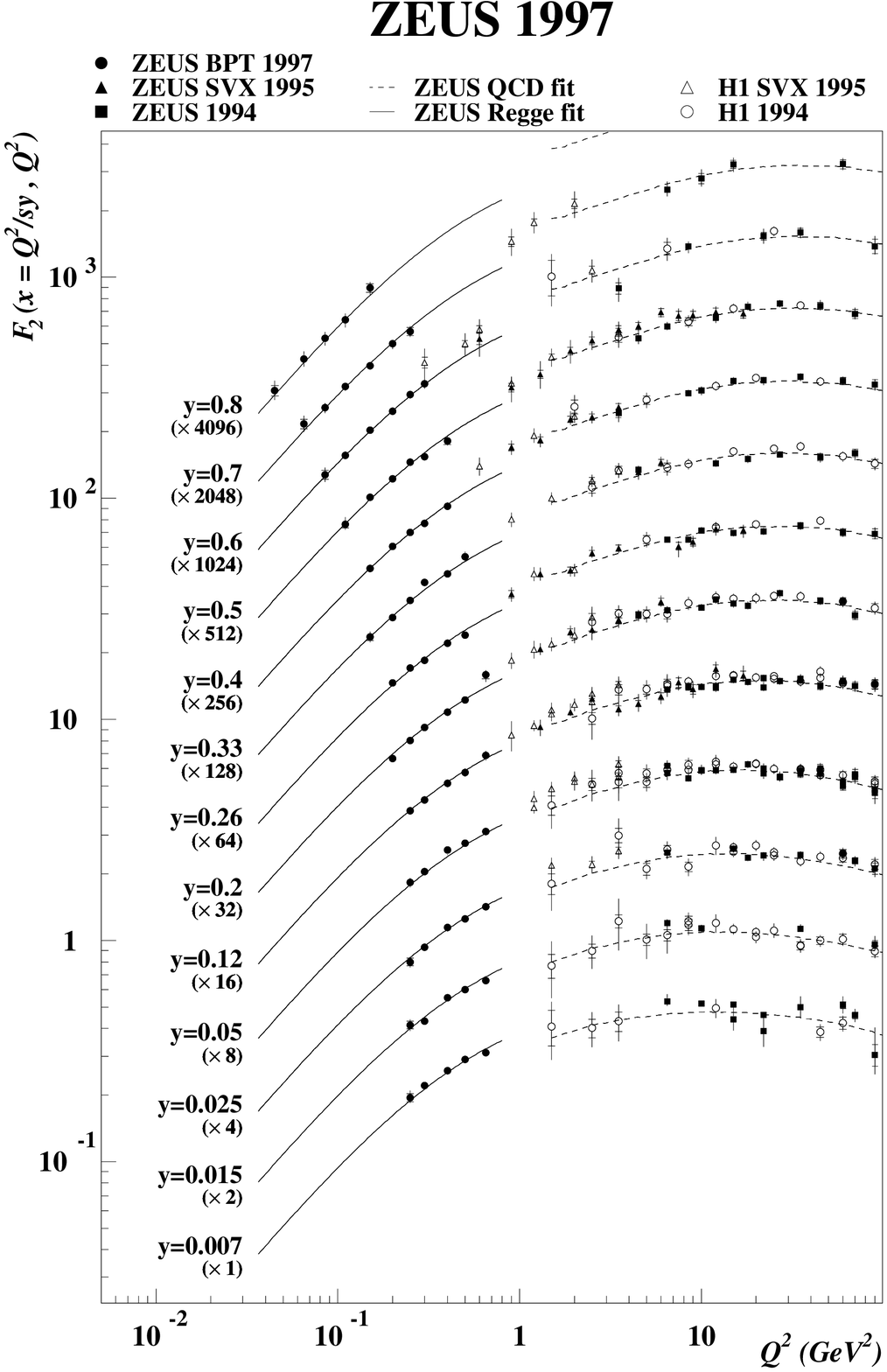,height=13cm}
\caption{
The proton structure function $F_2$ as function of $Q^2$ for fixed
values of $y$.
}
\label{fig:f2-lowq2}
\end{minipage}
\hspace*{2mm}
\begin{minipage}{7.2cm}
\psfig{figure=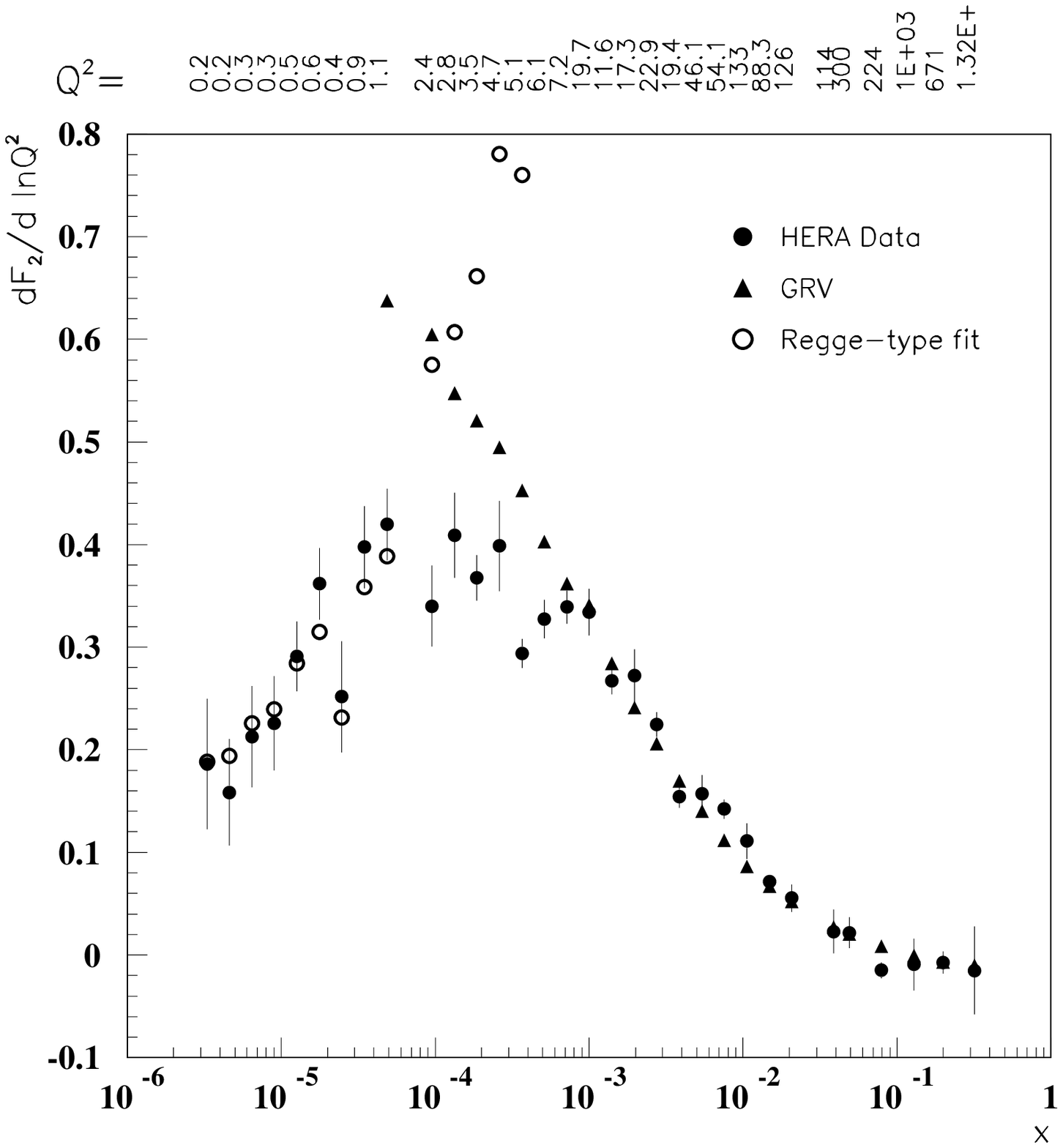,height=8cm}
\caption{
The slope of $F_2$ with respect to $Q^2$ as function of $x$ at given
$Q^2$ values.
}
\label{fig:caldwell}
\end{minipage}
\end{figure}

What is this strange behaviour of the gluons at low $x$ and low $Q^2$
tell us? If anything 'strange' is happening, the gluons will feel it
first because of the 9/4 color factor:
$\sigma_{ggp}=(9/4)\sigma_{q\bar{q}p}$.  One can calculate the
probability $P^D_g$ that an interaction on the gluon is associated
with diffraction. For a value of $Q^2$=4.5GeV$^2$ and $x<10^{-3}$ this
comes out to be $P^D_g \simeq$0.4~\cite{fs}, which is very close to
the black disc limit of 0.5. Does this mean that we see some kind of
saturation (unitarity effects)?  Such saturation effects could lead to
the suppression of 'large' $q\bar{q}$ configuration and lead to the
breakdown of DGLAP evolution in this region.  This can in fact be
nicely observed when $F_2$ is plotted as function of $Q^2$ for fixed
$y$~\cite{f2z}, displayed in figure~\ref{fig:f2-lowq2}, where one sees
the approximate scaling down to $Q^2 \sim$ 2 GeV$^2$ and then the
$1/Q^2$ decrease of $F_2$ towards the low $Q^2$ region.  This
behaviour, and also that of the Caldwell plot~\cite{rmp}, shown in
figure~\ref{fig:caldwell}, are well described by the saturation model
of Golec-Biernat and Wuesthoff~\cite{kgb}.

This section can be summarized by a quotation from Al Mueller's
message: {\it The turnover at small $Q^2$ (and small $x$) along with
the success of the Golec-Biernat Wuesthoff model in describing that
turnover (along with diffractive scattering) suggest that unitarity
limits have been reached when $Q^2 \approx$ 1 - 2 {\rm GeV$^2$} in the 
low $x$ region at HERA. In parton language this is saturation in the
small $x$ wave function of the proton.}

\newpage

\section{The photon - probing the vacuum}

HERA, the machine designed to teach us about the proton, turned out to 
be a good tool for studying the structure of both the real and the
virtual photon. 

A clear two component structure of the real photon has been observed
at HERA~\cite{photon}. The gluon distribution in the photon has been
extracted and also shows a rise with decreasing
$x$~\cite{h1gluong}. HERA thus provides an alternative way to that of
the $e^+e^-$ collider to measure the photon structure function and to
extract parton distributions in the photon.

Studies of these type in the DIS regime showed that also a virtual
photon can have a two component structure at low $x$ even at
virtualities as high as $Q^2 \sim$ 50 GeV$^2$~\cite{virt}. This of
course raises the question: is the virtual photon probing the
structure of the proton or are the partons from the proton probing the
structure of the photon? Who is probing whom? In fact, in the proton
rest frame $x$ can be given a spatial meaning: the coherence length
can be written as $l_c=1/(2m_px)\approx 0.1{\rm fm}/x$ which in the low $x$
region can be very large, much larger than the radius of the
proton. Thus at low $x$ we are actually not studying the structure of
the proton. To put it in Bjorken's words: {\it The important spacetime
regions are neither in the valence region of the hadron nor in the
``valence'' region of the photon probed.} At low $x$ we are probing
the structure of the vacuum.

\section{Very large $Q^2$ region}

The subject of the very large $Q^2$ region~\cite{highq2} actually
belongs to the future legacy of HERA, which the HERA upgrade will
produce. So far it produced a 'text-book' result showing that the
cross section of NC and CC reactions meet at $Q^2 \approx
M_Z^2$. Recently, the higher statistics data also showed clearly the
effect of the electroweak interference and the first measurements of
$xF_3$ were obtained. Searches for deviation from the SM caused
excitement.
Limits on exotic processes were obtained and await higher luminosity
data.

Our discussion in the previous section clearly indicates that in order 
to study the interior of the proton, one needs not only to go to
higher values of $Q^2$ but also to $x$ values in excess of 0.1. The
HERA upgrade should provide the luminosity necessary for obtaining
high enough statistics in the high $Q^2$ high $x$ region.

\section{Discussion}

The sharp rise of $F_2$ with decreasing $x$, the diffraction in DIS,
should we have been surprised? Some say yes, some say - in retrospect-
no. The fact, however, is that we were surprised and unprepared. The
original design of the detectors was not really well suited for the
low $x$, low $Q^2$ physics.  None of the DIS MC generators had
diffractive processes. We also took seriously parton distributions
which predicted a flat $F_2$ at low $x$.

To summarize the lesson learned from the low $x$ physics at HERA, we
have observed a tremendous progress in understanding of QCD dynamics
for high energy interactions and we might be faced with the existence
of a new QCD regime with high parton densities.

Another legacy of HERA was pointed out by James Stirling:
{\it Studying QCD at HERA has given us
great confidence that we can very precisely predict LHC cross
sections. Indeed, it is under active consideration to use Standard
Model $Z$ cross sections to measure the LHC luminosity, and it is only 
because the proton structure is pinned down so well by HERA that we
can trust the theoretical calculations.}

HERA has actually elevated QCD almost to the status of QED, in the
sense that QED processes are used to measure luminosities at HERA.
Not that there were doubts in the correctness of QCD. However, 
QCD is a complicated theory, and thus we can say that HERA 
in the year 2000 is a glorious triumph of minds dealing with QCD.

Did we however get a better picture of the proton? Rutherford's
experiment allowed him to say that most of the atom is empty with a
nucleus at its center. How do we see the proton? Some describe the
proton as being a 'Thompson' like proton: 'soup' of gluons filled with
pointlike quark 'raisins'. Other see it as a 'Rutherford' like proton:
3 centers of pointlike parton clouds. Asking Lonya Frankfurt, his
reply was that the outer 20\% of the proton is a pion cloud, while
nothing is known about the inner 80\%. Maria Krawczyk brought to my
attention the view of Altarelli, Cabbibo, Maiani and
Petronzio~\cite{acmp},  who
view the proton as 3 constituent quarks, each of which is a complex
object made out of pointlike partons. Where are all these partons
located in the proton? It would be nice to be able to stand in front
of a class and tell the students, just as Rutherford did about the
atom, how the proton looks like. The HERA upgrade program will
hopefully give us the answer.

\section*{Acknowledgments}
I would like to thank the organizers of this workshop, headed by John
Dainton and Josephine Zilberkweit, who did a marvelous job. Thanks
also for the Lord Mayor of Liverpool, Councillor Edwin Clein, who
promised us that in Liverpool all our dreams come true, and to the
Member of Parliament for the Liverpool Riverside Constituency,
Mrs. Louise Ellman, for her understanding and support of basic
research.

The discussion and input of the following people to this talk is
highly appreciated: H.~Abra\-mowicz, G.~Altarelli, J.~Bjorken,
A.~Caldwell, J.~Dainton, M.~Derrick, R.~Devenish, R.~Eichler,
F.~Eisele, L.~Frankfurt, D.~Haidt, R.~Klanner, M.~Krawczyk,
P.~Landshoff, E.~Levin, R.~Nania, J.~Stirling, M.~Strikman, A.~Wagner,
G.~Wolf, A.~Wroblewski.

This work was partially supported by the German-Israel Foundation
(GIF), by the US-Israel Binational Foundation (BSF), and by the Israel 
Science Foundation (ISF).

\newpage

\section*{Appendix}

\subsection*{DIS Kinematics}

\begin{figure}[hbt]
\begin{center}
\psfig{figure=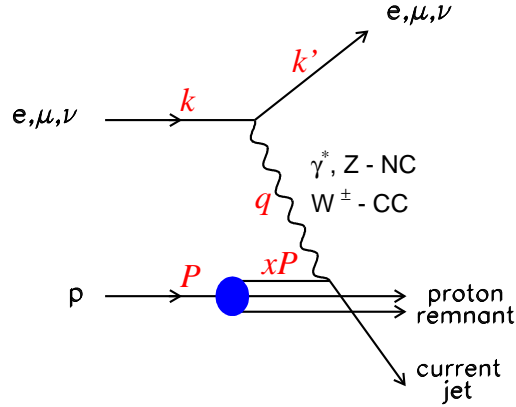,height=8cm}
\end{center}
\vspace{-1.5cm}
\caption {
Diagram describing a DIS process on a proton.
}
\label{fig:kinematics}
\end{figure}

A diagram describing a DIS process on a proton is shown in
figure~\ref{fig:kinematics}. A lepton with mass $m_l$ and four-vector
$k(E_l,\vec{k})$ interacts with a proton with mass $m_p$ and
four-vector $P(E_p,\vec{p})$ through the exchange of a gauge vector
boson, which can be $\gamma$, $Z^0$ or $W^\pm$, depending on the
circumstances. The four-vector of the exchanged boson is
$q(q_0,\vec{q})$.

With these notations one can define the following variables,
\begin{eqnarray}
         q    &=&     k - k^\prime     \\
       \nu  &\equiv&  \frac{P\cdot q}{m_p}      \\
         y  &\equiv&  \frac{P\cdot q}{P\cdot k} \\
       W^2    &=&     (P + q)^2                 \\
         s    &=&     (P + k)^2.
\end{eqnarray}
The meaning of the variables $\nu$ and $y$ is most easily realized in
the rest frame of the proton. In that frame $\nu$ is the energy of the
exchanged boson, and $y$ is the fraction of the incoming lepton energy
carried by the exchanged boson.  The variable $W^2$ is the squared
center of mass energy of the gauge--boson proton system, and thus also
the squared invariant mass of the hadronic final state. The variable
$s$ is the squared center of mass energy of the lepton proton system.

The four momentum transfer squared at the lepton vertex can be approximated
as follows (for $m_l, m_l^\prime \ll E, E^\prime$),
\begin{equation}
        q^2 = (k-k^\prime)^2
            = m_l^2+{m_l^\prime}^2-2kk^\prime
            \approx -2EE^\prime (1-\cos\theta)
            < 0 \ .
\label{eq:q2}
\end{equation}
The scattering angle $\theta$ of the outgoing lepton is defined
with respect to the incoming lepton direction.
The variable which is mostly used in DIS is the negative value of the four
momentum transfer squared at the lepton vertex,
\begin{equation}
                        Q^2\equiv-q^2 \ .
\end{equation}
One is now ready to define the other variable most frequently used in
DIS, namely the dimensionless scaling variable $x$,
\begin{equation}
                   x \equiv \frac{Q^2}{2 P\cdot q} \ .
\end{equation}
To understand the physical meaning of this variable, one goes to a
frame in which masses and transverse momenta can be neglected - the
so-called infinite momentum frame. In this frame the variable $x$ is
the fraction of the proton momentum carried by the massless parton
which absorbs the exchanged boson in the DIS interaction. This
variable, defined by Bjorken, is duly referred to as Bjorken-$x$.

The diagram in figure~\ref{fig:kinematics} describes both the
processes in which the outgoing lepton is the same as the incoming
one, which are called neutral current reactions (NC), as well as those
in which the nature of the lepton changes (conserving however lepton
number) and which are called charged current processes (CC). In the NC
DIS reaction, the exchanged boson can be either a virtual photon
$\gamma^*$, if $Q^2$ is not very large and then the reaction is
dominantly electromagnetic, or can be a $Z^0$ which dominates the
reaction at high enough $Q^2$ values and the process is dominated by
weak forces. In case of the CC DIS reactions, only the weak forces are
present and the exchange bosons are the $W^\pm$.

\section*{Kinematics of diffractive scattering}

The variables used to analyze diffractive scattering are introduced
for $ep$ DIS.

\begin{figure}[htb]
\begin{center}
\epsfig{file=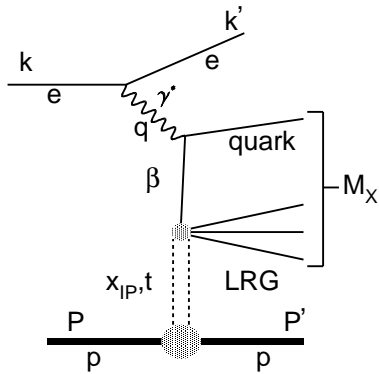,height=5cm}
\caption{Schematic diagram for diffractive DIS in $ep$ interactions.} 
\label{fig:dis-diag}
\end{center}
\end{figure}
A diagram for diffractive scattering in
DIS, where the diffracted state is separated from the scattered proton
by a large rapidity gap (LRG), is presented in
figure~\ref{fig:dis-diag} and all the relevant four vectors are
defined therein. 
In addition to the usual DIS variables, defined above,
the variables used to describe the diffractive final state are,
\begin{eqnarray}
t &=& (P-P^\prime)^2 \, ,
\label{eq:deft} \\
\xpom &=& \frac{q\cdot (P-P^\prime)}{q \cdot P}
\simeq \frac{M_X^2+Q^2}{W^2+Q^2} \, ,
\label{eq:defxpom} \\
\beta &=& \frac{Q^2}{2q \cdot (P-P^\prime)} = \frac{x}{\xpom}
\simeq \frac{Q^2}{Q^2+M_X^2} \, .
\label{eq:defbeta}
\end{eqnarray}
$\xpom$ is the fractional proton momentum which
participates in the interaction with $\gamma^*$. It is sometimes denoted by
$\xi$. $\beta$ is the equivalent of Bjorken $x$ but
relative to the exchanged object. $M_X$ is the invariant mass of the
hadronic final state recoiling against the leading proton,
$M_X^2=(q+P-P^\prime)^2$.  The approximate relations hold for small
values of the four-momentum transfer squared $t$ and large $W$,
typical of high energy diffraction.


To describe diffractive DIS, it is customary to choose the variables
$\xpom$ and $t$ in addition to the usual $x$ and $Q^2$ in the cross
section formula. The diffractive contribution to $F_2$ is denoted by
$F_2^D$ and the corresponding
differential contributions are
\begin{equation}
F_2^{D(3)}=\frac{dF_2^D}{d\xpom} \, , \ \ \ \ \ 
F_2^{D(4)}=\frac{dF_2^D}{d\xpom dt} \, . 
\end{equation}
The contribution from the longitudinal structure function is omitted
for simplicity.

The four-fold differential cross section for $ep$ scattering can be
written as
\begin{equation}
\frac{d^4\sigma^D_{ep}}{ d\,\xpom d\,t d\,x d\,Q^2 }
=\frac{2\pi \alpha^2}{x Q^4} \left[ 1+(1-y)^2\right] 
F_2^{D(4)}(x,Q^2,\xpom,t) \, . \label{eq:f2d4} 
\end{equation}

The structure function $F_2$ is related to the absorption cross
section of a virtual photon by the proton, $\sigma_{\gamma^\star p}$.
For diffractive scattering, in the limit of high $W$ (low $x$),
\begin{equation}
F_2^{D(4)}(x,Q^2,\xpom,t) = \frac{Q^2}{4\pi^2\alpha}
\frac{d^2\sigma^D_{\gamma^\star p}}{ d\,\xpom d\,t} \, .
\label{eq:gstarp}
\end{equation}

\end{document}